\documentclass{icrc29}
\usepackage{graphicx,amssymb,amsmath,times}
\begin{document}

\title{Performance of the Pierre Auger Observatory Surface Array}

\author{The Pierre Auger Collaboration}

\presenter{Presenter: X. Bertou (bertou@auger.org.ar), \
arg-bertou-X-abs1-he14-oral}

\maketitle

\begin{abstract}
The surface detector of the Pierre Auger Observatory is a 1600 water Cherenkov
tank array on a triangular 1.5\,km grid. The signals from each tank are read
out using three 9'' photomultipliers and processed at a sampling frequency of
40\,MHz, from which a local digital  trigger efficiently selects shower
candidates. GPS signals are used for time synchronization and a wireless
communication system connects all tanks to the central data acquisition system.
Power is provided by a stand-alone solar panel system.\\ 
With large ambient temperature variations, that can reach over 20 degrees in
24 hours, high salinity, dusty air, high humidity inside the tank, and
remoteness of access, the performance and reliability of the array is a
challenge. Several key parameters are constantly monitored to ensure consistent
operation.\\ 
The Surface Array has currently over 750 detectors and has been in reliable operation
since January 2004. Good uniformity in the response of different detectors
and good long term stability is observed.
\end{abstract}

\section{Introduction}
\label{intro}

The Pierre Auger Observatory aims at unveiling the secrets of Ultra High Energy
Cosmic Rays (UHECR) through the observation of the Extensive Air Showers (EAS)
they produce in the atmosphere. It combines four fluorescence detector (FD)
sites with a ground array to achieve hybrid detection of the EAS\,\cite{hybrid}. The
fluorescence detectors, however, operate only at night, with limited moon light
contamination, affecting the statistics of the hybrid data set. To achieve high
exposure, the surface array, with an uptime close to 100\,\%, is fundamental.

An Auger Surface Detector (SD) station is a 10\,m$^2$ base, 1.5\,m tall cylindrical
rotomolded plastic tank. It encloses a Tyvek\textsuperscript{\textregistered} liner filled with
locally produced purified water (resistivity: 15\,M$\Omega$\,cm) up to a 1.2\,m
level. Three 9'' Photonis XP1805 photomultiplier tubes
are used to collect the Cherenkov light emitted by particles
crossing the tank. Signal is extracted both from the anode and the last dynode,
the latter being amplified to achieve a larger final dynamic range
extending from a few to about $10^5$ photoelectrons. All
channels are connected to a 5 pole anti-aliasing filter with a cutoff at
20\,MHz, digitized at 40\,MHz by 10 bit FADC, a digital trigger is
operated
by a local CPU. Timing is obtained by a GPS Motorola unit, and communication to
the Central Data Acquisition System (CDAS) is done via a custom built wireless
communication system. Two solar panels charging two 12\,V batteries
provide the 10\,W used by the electronics.
Each detector is therefore independent and can start operating upon
installation, independently of other detectors in the array.
More details about the SD can be found in \cite{nim}.

\begin{figure}
\begin{minipage}[t]{0.48\textwidth}
\mbox{}
\begin{center}
\includegraphics*[width=0.8\textwidth,height=0.64\textwidth,angle=0,clip]{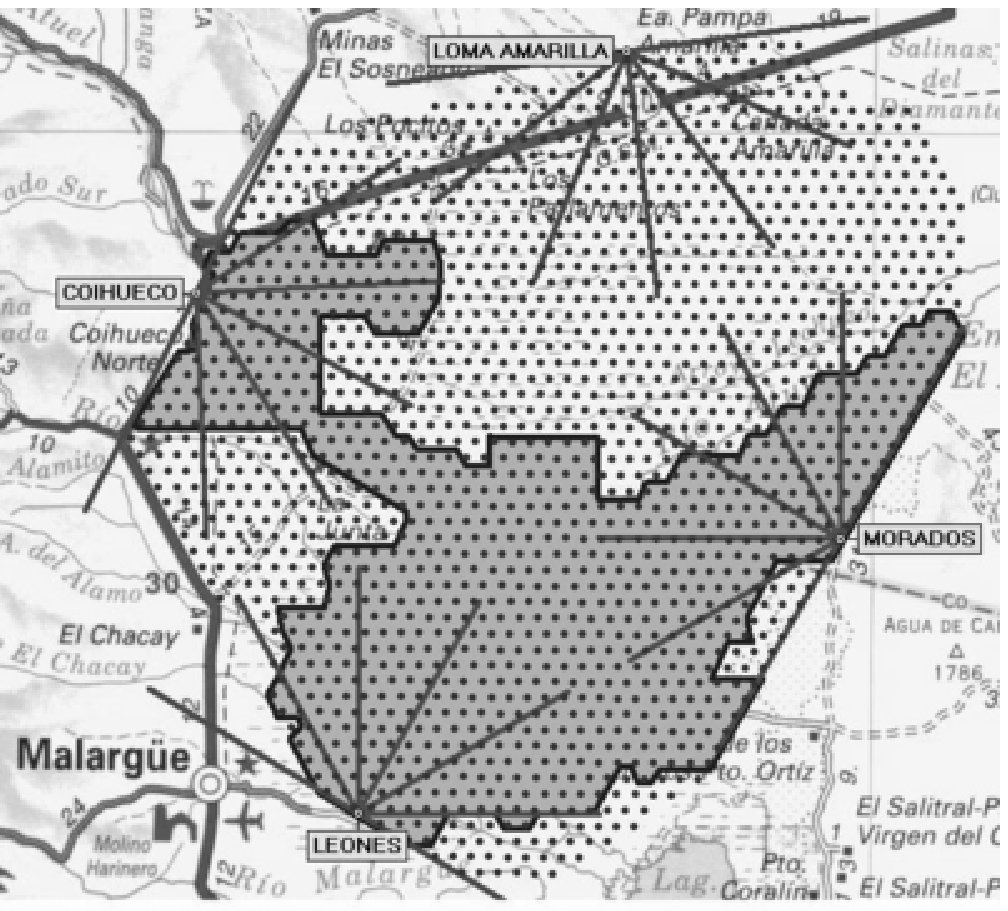}
\caption{SD Array status on June 2005. The map is roughly 70\,km by 50\,km,
close to the Malarg\"ue city (lower left corner). Each dot represents one
station, and the array is surrounded by the 4 FD buildings, displayed with the
field of view of their 6 bays. The hatched region corresponds to the
currently
deployed stations, one half of the final 1600 stations array. The shape was
driven by two main concerns: install stations close to the FD buildings to
improve the hybrid acceptance, and take advantage of dry summer to install them
in areas difficult to access.
\label{fig:array}}
\end{center} 
\end{minipage}
\hfill
\begin{minipage}[t]{0.48\textwidth}
\mbox{}
\begin{center}
\includegraphics*[width=0.9\textwidth,height=0.43\textwidth,angle=0,clip]{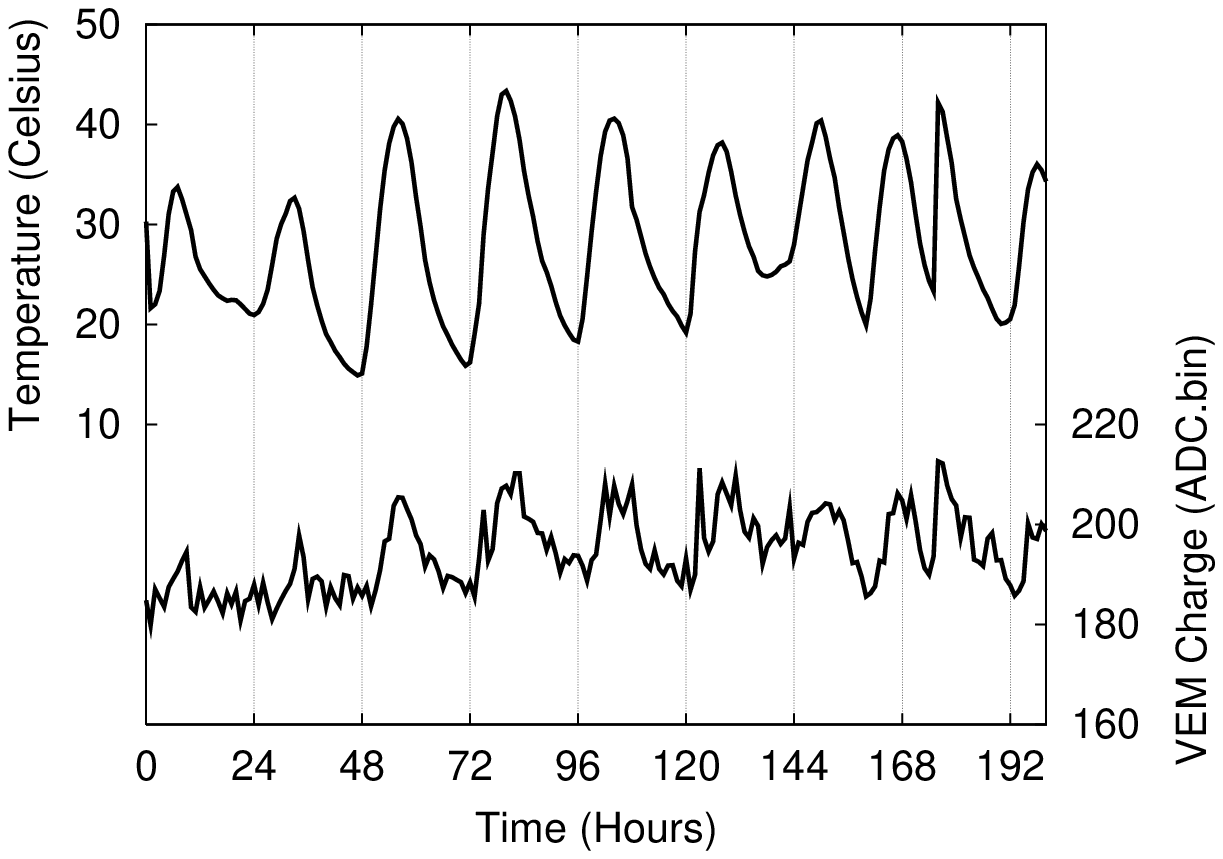}\\
\includegraphics*[width=0.9\textwidth,height=0.43\textwidth,angle=0,clip]{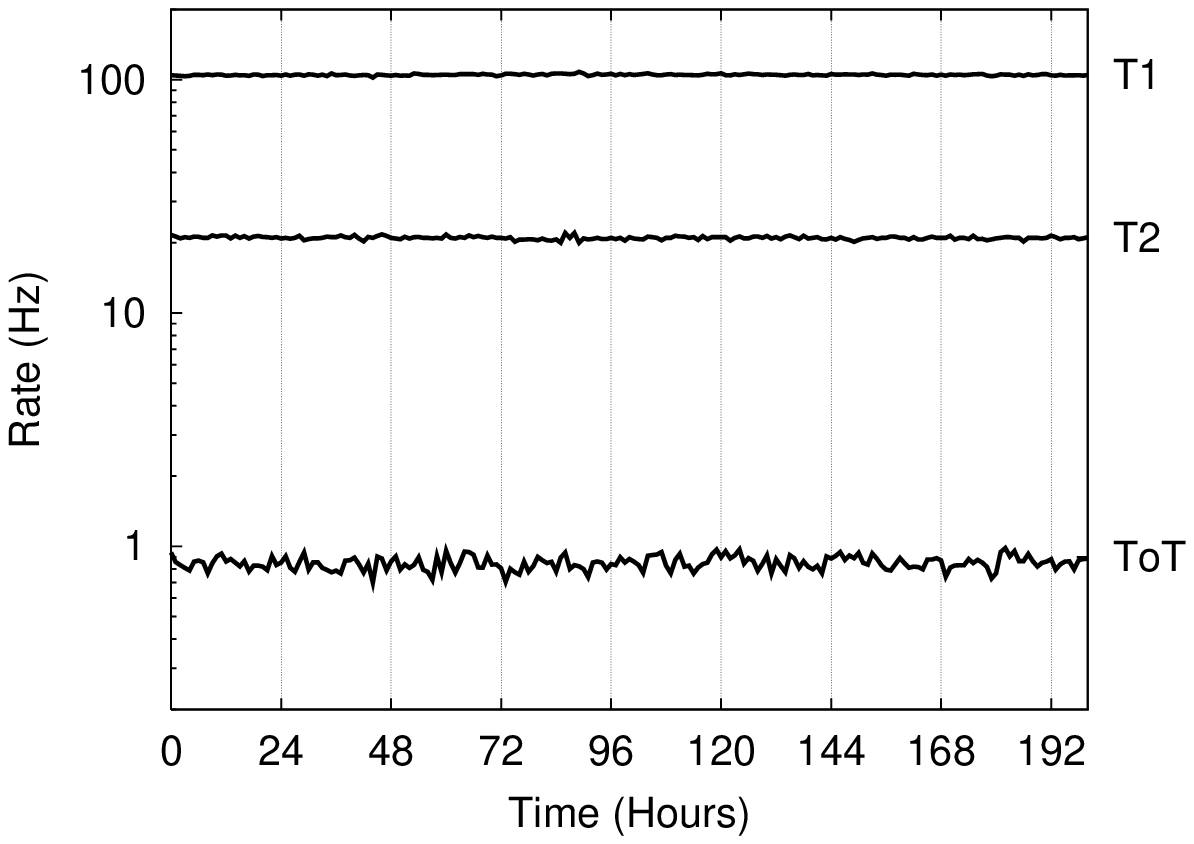}
\caption{
Top: evolution of the electronics temperature (top curve) and of the VEM
charge (bottom curve) of station 102 over a week in April 2005.  The
correlation is clear.  Bottom: trigger rates (T1: first level trigger, T2:
second level trigger, ToT: Time over threshold trigger rate\,\cite{trig}) for the
same station 102 over the same period.
\label{fig:vem}}
\end{center} 
\end{minipage}
\end{figure}

The Auger Surface Array will consist of 1600 of these detectors, on a
3000\,km$^2$ triangular grid of 1.5\,km spacing. Since January 2004, the array
has been in stable operation and has grown at a steady rate of about 9 tanks per week.
Figure \ref{fig:array} shows the status in June 2005.  Each tank is deployed and
its position is verified with differential GPS technique. Even if the landscape
sometimes forces some displacements from the perfect triangular geometry,
50\,\% of the tanks are at less than 5\,m from the theoretical position, and
90\,\% at less than 20\,m. The exact position is used to operate the GPS
in position hold mode, achieving better than 20\,ns time resolution.

\section{Environment monitoring and Calibration stability}

The environment to which an Auger Surface Detector is exposed is somewhat
hostile for the electronics. At 1400\,m a.s.l. and with clear skies, day-night
temperature variations are of the order of 20$^\circ$C. The terrain
properties vary on the 3000\,km$^2$ of the array, and some detectors have to
face high salinity corrosive air, while others face dusty air from sandy
ground. Humidity is present both in the ground (rivers, groundwater) and in the
tank due to condensation.

To monitor the whole array accurately various sensors are installed in
each tank. This information is sent to the CDAS
every 6 minutes.  Temperature is measured on each PMT base, on the
electronics board, and on each battery. PMT voltage and current are also
monitored, as well as solar panel voltages, individual battery voltage, and
charge current. These data are used to detect a wide range of
failures, from broken solar panels to discharging batteries, and
correlations like unstable PMT behavior related to temperature.

Weather stations reporting temperature, pressure, humidity, wind speed and
direction are installed at each fluorescence site and in the center of the
array, to complete the environmental monitoring. These data allow extra checks
such as the influence of the pressure on the calibration.

The calibration is operated online every minute\,\cite{calib}, and sent to CDAS
every 6 minutes for monitoring. A number of quantities are available to check
the behavior of the tank: baseline values, single muon peak signal,
single muon average charge, dynode/anode ratio, and PMT stability. In addition,
high statistics histograms of the charge distribution from muons (150\,000 counts) and average pulse shapes are
sent every time a station records an event.

This information allows detailed monitoring of the evolution of the calibration
and various cross-checks. Figure \ref{fig:vem} shows the evolution of the VEM
(Vertical Equivalent Muon, average charge deposited by a vertical and central
through going muon) and its correlation with temperature for a specific detector.
The online calibration takes into account most temperature fluctuations
at the local trigger level as reflected in the same figure by the stability of
the trigger rate over time\,\cite{trig}. Over the whole array, correlation of
the trigger rate with temperature are -0.04$\pm$0.03\,\% per degree for
first trigger level (T1), 0.08$\pm$0.05\,\% per degree for the second
level trigger (T2), and 0.2$\pm$0.5\,\% per degree for the Time over
Threshold trigger (ToT). The SD array therefore
operates with stable trigger threshold even with 20 degrees daily
temperature variations. It is clear from the plots of figure \ref{fig:vem} that
calibrating the array only once per day would introduce a significant bias in
the trigger threshold, of the order of 10\,\%.

\begin{table}[!hb]
\begin{center}
\begin{tabular}{|c|c|c|}\hline
issue & fraction of stations \\
\hline
\hline
PMT connector & 0.9\,\% \\
PMT unstable & 2.5\,\% \\
PMT various & 1.4\,\% \\
Electronics generic & 4\,\% \\
Power issue & 5\,\% \\
\hline
Baseline & 0.3\,\%\\
VEM & 5\,\%\\
Charge/Peak & 0.9\,\%\\
Dynode/Anode & 0.5\,\% \\
\hline
\end{tabular}
\quad\quad
\begin{tabular}{|c|c|}\hline
Cause & Dead time \\
\hline
\hline
Power cut & 0.9\,\% \\
Communication shortout & 0.9\,\% \\
Acquisition deadtime & 1.1\,\% \\
Software Updates & 0.6\,\% \\
Individual detector failure& 2.2\,\% \\
\hline
\hfill Total& 5.7\,\% \\
\hline
\hline
\hfill On-time & 94.3\,\% \\
\hline
\end{tabular}
\end{center}
\caption{
Left: fraction of poorly performing detectors. First part of the table describes
electronics related issues, and second part its impact on calibration values
(fraction of stations with the calibration parameter outside of established
range or with fluctuations above 5\,\%).
\label{tab:sde}
\label{tab:dead}
Right: dead times of the SD array in 2004. See text for details.
}
\end{table}

Finally, analysis of all these parameters allows to detect poorly
performing
detectors. Table \ref{tab:sde} shows the fraction of detector in which some
monitoring parameters fall outside of the expected range or exhibit large
fluctuations. While this fraction is already small, it is important to remember that
most of the detectors have been recently deployed and infancy problems are the
major cause of improper functioning. As a reference, out of 96 PMT issues, only
2 manifested problems after 18 months of correct behavior. The remaining
94 showed problems in the first days of operation.

\section{Central Data Acquisition}

The last step monitored to ensure the quality of the Auger SD data set
is done at the system acquisition level. The second level trigger rate for each
station are registered every second allowing a precise knowledge of the dead time of
the detectors.

The acquisition is fully automated and no operator is needed for data taking.
A ``watchdog'' restarts the system in case of failure and an UPS guarantees
operation in case of power cut. Information from the CDAS processes are kept to
diagnose possible crashes.

Simple quantities such as the number of stations in operation and the event rate,
and more complex ones such as the rate of physics events are checked daily to
validate the data acquisition period. Over 2004, the total on-time of the system
has been $\sim$\,94\,\%, including all kinds of dead time (individual detectors down,
general power cuts, software upgrades, etc.). Table \ref{tab:dead} summarizes
the various contributions. It should be noted that this on-time was obtained
while priority was being given to the building of the Observatory (deploying new
detectors) over its operation (repairing failing ones), and with evolving
software for the detectors, the communication system, and the CDAS. It seems
clear that once the Observatory is completed and the software stabilized,
an on-time above 95\,\% should be achievable.

\begin{figure}
\begin{center}
\includegraphics*[width=0.45\textwidth,height=0.2\textwidth,angle=0,clip]{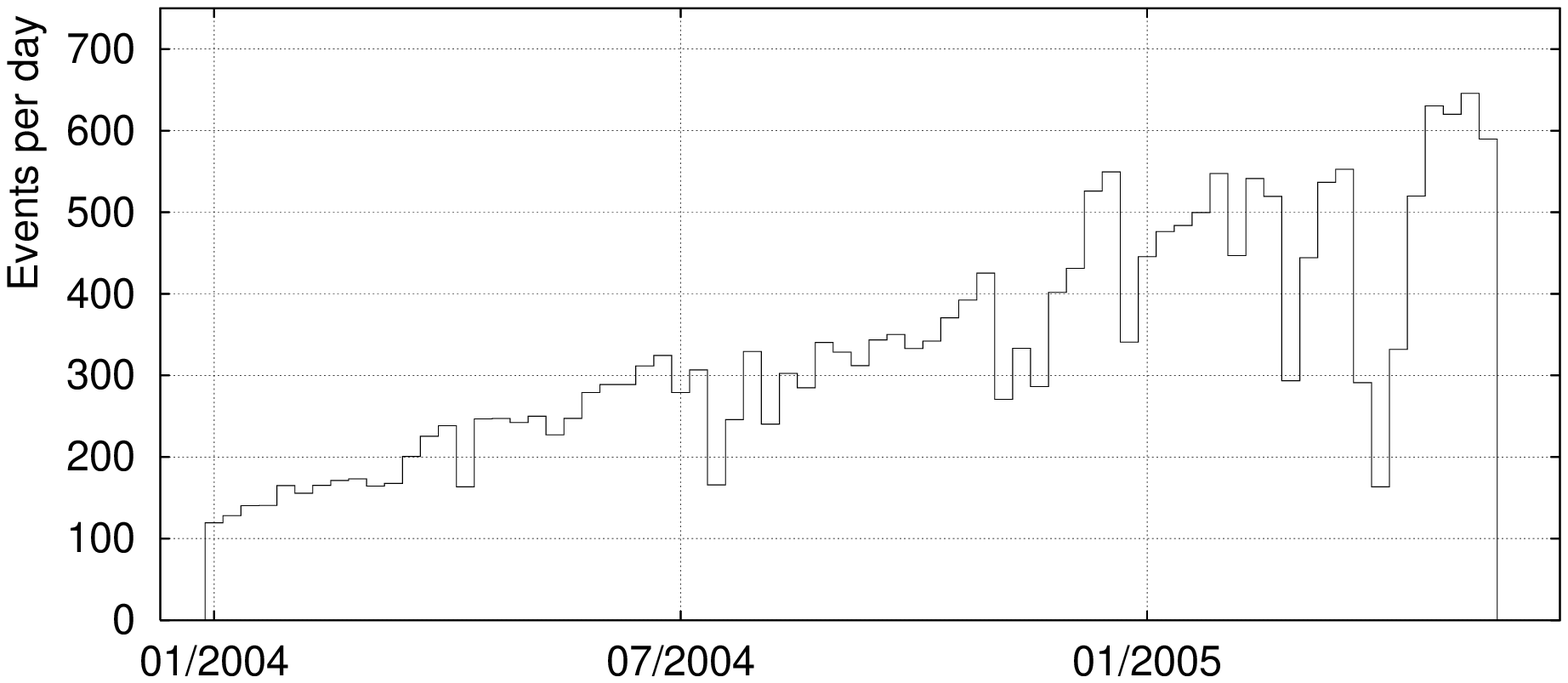}
\includegraphics*[width=0.45\textwidth,height=0.2\textwidth,angle=0,clip]{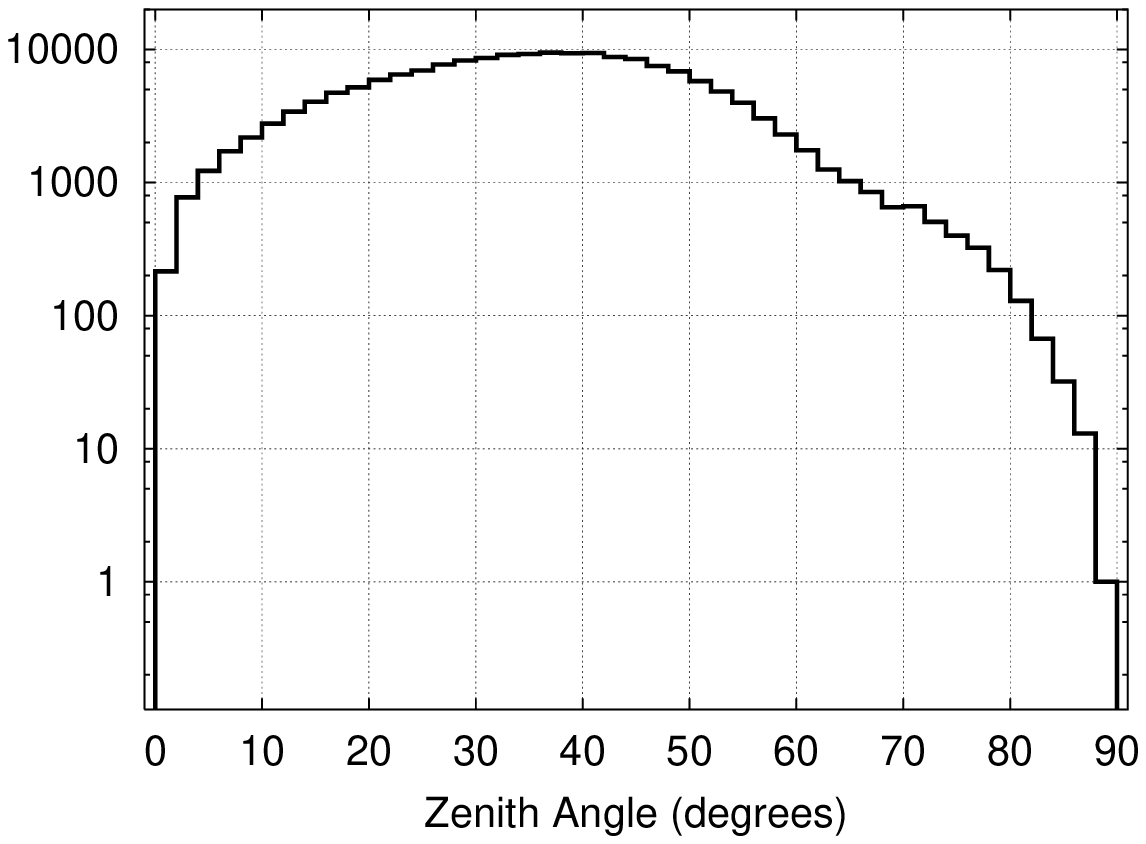}
\caption{
Left: Evolution of event rate with time.
An average event rate of about one physics event per day per station is
observed. The consequence of a major software upgrade in April 2005 is
visible as a dip in the plot.
\label{fig:ev}
Right: Zenith angle distribution for all events
observed up to June 2005. Large zenith angle events manifest themselves
as high multiplicity (20 tanks or more) due to the compact aspect of the
array when viewed at large zenith angle.
\label{fig:ev2}}
\end{center} 
\end{figure} 

Figure \ref{fig:ev} shows the evolution of the physics event 
rate as a function of time. It is roughly related to
the number of active stations by 0.9 event per day per station.
The zenith angle distribution of these events is also plotted. A significant number
of very horizontal events is observed\,\cite{hor}.

\section{Conclusion}

The surface array of the Pierre Auger Observatory has been in 
operation for more than a year and has shown very stable behavior. On-time for
2004 is above 94\,\%.  Detectors have been shown to be able to compensate
important daily temperature fluctuations. A small number of poorly
performing stations have been identified, and most failures appear in the
first days of operation. Fraction of stable detectors is increasing and
reached 97\,\% at the time of writing.

Up to June 2005, more than 180\,000 events were recorded with an average
rate of about 0.9 per station per day. Once the array is completed, a
rate of about 1500 physics events per day is expected. A significant
number of very horizontal events are detected, offering a novel view of
EAS.  For events above 10$^{19}$\,eV, reconstruction accuracy is around 1.4 degree for
direction\,\cite{ang} and around 10\% for the SD energy estimator (signal at 1000\,m
from the core\,\cite{s1000}).

The SD array has shown its potential in the period from January 2004 to
June 2005, achieving below 60 degrees of zenith angle a total exposure of
1750\,km$^2$\,sr\,yr.  It will reach its full size, and gain one order of
magnitude exposure within the next two years, hopefully unveiling the secrets
of the UHECR.

\end{document}